\documentclass[reprint,superscriptaddress,amsmath,amssymb,a4paperaps,pra,nofootinbib]{revtex4-1}
\usepackage{dcolumn,amssymb,amsmath,amsfonts,graphicx,latexsym,float,xcolor}
\usepackage{epstopdf}
\usepackage{nameref}
\usepackage{txfonts}
\usepackage{ulem}
\begin{document}

\title{Dynamical formation of two-fold fragmented many-body state\\ induced by an impurity in a double-well}

\author{Jie Chen}
\email {jie.chen@physnet.uni-hamburg.de}
\affiliation{Department of Physics, Center for Optical Quantum Technologies, University of Hamburg, Luruper Chaussee 149, 22761 Hamburg, Germany}
\author{Simeon I. Mistakidis}
\email{symeon.mystakidis@cfa.harvard.edu}
\affiliation{ITAMP, Center for Astrophysics $|$ Harvard $\&$ Smithsonian, Cambridge, MA 02138 USA}
\affiliation{Department of Physics, Harvard University, Cambridge, Massachusetts 02138, USA}
\author{Peter Schmelcher}
\email{pschmelc@physnet.uni-hamburg.de}
\affiliation{Department of Physics, Center for Optical Quantum Technologies, University of Hamburg, Luruper Chaussee 149, 22761 Hamburg, Germany}
\affiliation{The Hamburg Center for Ultrafast Imaging, University of Hamburg, Luruper Chaussee 149, 22761 Hamburg, Germany}
\date{\today}

\begin{abstract}
We unravel the correlated quantum quench dynamics of a single impurity immersed in a bosonic environment confined in an one-dimensional double-well potential.  A particular emphasis is placed on the structure of the time-evolved many-body wave function by relying on a Schmidt decomposition whose coefficients directly quantify the number of configurations that are macroscopically populated. For a non-interacting bosonic bath and weak postquench impurity-bath interactions, we observe the dynamical formation of a two-fold fragmented many-body state which is related to  intra-band excitation processes of the impurity and manifests as a two-body phase separation (clustering) between the two species for repulsive (attractive) interactions. Increasing the postquench impurity-bath coupling strength leads to the destruction of the two-fold fragmentation since the impurity undergoes additional inter-band excitation dynamics. By contrast, a weakly interacting bath suppresses  excitations of the bath particles and consequently the system attains a weakly fragmented many-body state. Our results explicate the interplay of intra- and inter-band impurity excitations for the dynamical generation of fragmented many-body states in multi-well traps and for designing specific entangled impurity states.
\end{abstract}

\maketitle

\section{Introduction}

The dynamics of 
isolated 
cold atom many-body (MB) systems 
is a topic of vigorous theoretical interest~\cite{Cazalilla,Sowinski,review1D_Mistakidis}. 
It can be readily assessed in current experiments offering   
high degree of controllability~\cite{cold_atom_rev} e.g., with respect to the 
atom number~\cite{Wenz,Zurn} or the interatomic interactions \cite{Feshbach_1, Feshbach_2, Feshbach_3, BH_exp_1,BH_exp_2, BH_exp_3}. 
Furthermore, restricting the  
atomic motion into only a few or even 
solely two single-particle modes can significantly simplify the quantum simulation of MB systems 
while still retaining the possibility to unravel emergent 
quantum MB phenomena. 
A relevant widely used setup is the bosonic Josephson junction 
emulated through a 
MB bosonic gas in a one-dimensional (1D) double-well (DW) 
\cite{DW_exp_1,DW_exp_2, DW_exp_3} 
and sharing analogies with the Josephson effect initially predicted for tunneling of Cooper pairs between two weakly linked superconductors~\cite{BJJ_1,BJJ_2}. 
In this context, several 
intriguing phenomena have been found 
such as Josephson oscillations \cite{BJJ_Rabi_1, BJJ_Rabi_2, BJJ_Rabi_3}, macroscopic quantum self-trapping 
\cite{DW_exp_3, BJJ_Rabi_1, BJJ_Rabi_2}, 
formation of an atomic squeezed state~\cite{BJJ_Squeeze_1, BJJ_Squeeze_2} and strongly correlated tunneling processes in few-body systems~\cite{BJJ_Few_1, BJJ_Few_2, BJJ_Few_3, BJJ_Few_4, BJJ_Few_5}. 

Substantial theoretical and experimental efforts have been recently focusing on strongly particle imbalanced cold 
atomic mixtures 
for both bosonic~\cite{Bose_polaron_1, Bose_polaron_2} and fermionic~\cite{Fermi_polaron_1, Fermi_polaron_2, Fermi_polaron_3} settings where quasi-particle formation~\cite{polaron_conmat_1, polaron_conmat_2, polaron_conmat_3, polaron_conmat_4}, and in particular polarons~\cite{Massignan_rev,Schmidt_rev}, can take place. 
Despite the quasi-particle concept, impurity systems can be proven very useful for probing transport phenomena~\cite{Johnson_lat,Palzer_lat,Bruderer_lat,Theel_DW}, induced correlations~\cite{Dehkharghani_ind,ind_int_1, Mistakidis_ind, Mukherjee_ind}, facilitate bound state formation~\cite{Naidon,Alhyder,Will,Camacho_bipol,Brauneis_bound}, spontaneous generation of non-linear pattern formation~\cite{Yang_sol,Koutentakis_sol,Tajima_shock} as well as study relaxation processes~\cite{Lausch_relaxation,Boyanovsky_dissipative_pol,Mistakidis_dissipative_pol}. 
Overall, entanglement is supposed to be a crucial ingredient in these systems in part due to their few-body nature and also the non-negligible impurity-bath coupling. 
As such, it is particularly interesting to understand entanglement growth in these systems and its origin which can be traced back to the structure of the underlying MB wave function.   
In another context, quantum entanglement, a nonlocal property that is  inherent to quantum mechanics, plays an important role in quantum information proccesess and it is 
expected to provide the main resource of the quantum speed-up in quantum computation and  
communication~\cite{entanglement_1}. 

Stimulated by the above progress, studies on quantum entanglement lead to further insights into many- and few-body systems~\cite{entanglement_2,entanglement_3,entanglement_4,entanglement_5}. 
Characteristic examples include, for instance, the perturbative calculation of inter-species entanglement in Bose–Bose mixtures in optical lattices~\cite{entanglement_2} 
probing of quantum entanglement in a many-body localized phase~\cite{entanglement_3}, the production of entangled states in mesoscopic atomic systems~\cite{entanglement_4} as well as comparisons of different entanglement measurements in trapped few-particle systems~\cite{entanglement_5}. 
Note, however, that while most of these investigations focus on either the ground-state properties of the system or are limited to certain entropic measurements, studies exploring the creation of quantum entanglement and its relation 
to the structure of the associated MB wave function are still rare.

We investigate the non-equilibrium correlated quantum dynamics of a single impurity immersed in a non-interacting bosonic environment confined within a 1D DW potential~\cite{Impurity_BH_1, Impurity_BH_2, Impurity_BH_3, Impurity_BH_4, Impurity_BH_5, Impurity_BH_6}. 
To provide a complete description of the dynamical response of the binary mixture in the strongly correlated regime, we employ a numerically exact diagonalization approach which enables us to take all emergent correlations into account. 
A focus is placed on the microscopic configurations of the 
MB wave function following  
a quench of the impurity-bath interaction strength.  
Specifically, the time-dependent MB state is analyzed in terms of 
a Schmidt decomposition which provides access to the participating  
configurations, namely the Schmidt product states consisting of the impurity and the bath species as well as the degree of fragmentation through the respective coefficients. 

Considering weak postquench impurity-bath interactions, we demonstrate the \textit{dynamical formation of a two-fold fragmented MB state} in which two configurations are almost equally populated 
for long evolution times. 
The generation of the two-fold fragmented MB state is signified by the "collapse-and-revival" behavior of the first two Schmidt numbers and it is related to the impurity's  
intra-band excitation processes. 
This fragmentation mechanism is also imprinted in the impurity-bath two-body  
density distribution, manifesting as a  
two-body phase separation (clustering) among the species for repulsive (attractive) interactions. 
Furthermore, we explore  
the effect of either an increasing postquench interspecies coupling strength to strong interactions 
or a weakly interacting bath. 
It is showcased that for strong impurity-bath couplings, the impurity experiences inter-band excitations which leads to the substantial occupation of 
higher than the first two energetically lowest Schmidt states. 
These interband excitations are also visualized on the impurity's single-particle density featuring a spatially delocalized response and characterized by a multi-hump structure. 
In contrast, a weakly interacting bath suppresses 
excitations of the bath particles, while the impurity features a breathing motion and consequently the system attains a weakly fragmented MB state. 

This work is organized as follows. In Sec.~\ref{Setup}, we present the impurity setting under consideration and its tight-binding description. 
Sec.~\ref{quench_weak} elaborates on our main observation: the \textit{dynamical formation of a two-fold fragmented MB state} in the weakly impurity-bath coupling regime when the host is non-interacting. 
In Sec.~\ref{quench_strong}, we explicate the effect of strong impurity-bath postquench interactions on the impurity's fragmentation dynamics, while in Sec.~\ref{interacting_bath} we study the impact of finite intraspecies bath interactions on the MB dynamics of the mixture. 
Finally, our conclusions and outlook are provided in Sec.~\ref{Conclusions}.

\section{Tight-binding description of the impurity-bath setup}\label{Setup}

Our two-component setting consists of a single impurity $N_{I} = 1$ 
embedded in a gas of $N_{B} = 100$ bosons. 
The mass-balanced ($m_B=m_I$) mixture is trapped in a 1D symmetric DW 
and its MB Hamiltonian is given by $\hat{H} = \hat{H}_{I} + \hat{H}_{B} +  \hat{H}_{IB}$, where 
\begin{align}
\hat{H}_{I} &=\int dx~\hat{\psi}^{\dagger}_{I}(x) \textit{h}_{I}(x) \hat{\psi}_{I}(x), \nonumber\\
\hat{H}_{B} &=\int dx~\hat{\psi}^{\dagger}_{B}(x) \textit{h}_{B}(x) \hat{\psi}_{B}(x)   \nonumber\\
&+\frac{g_{BB}}{2}\int dx~\hat{\psi}^{\dagger}_{B}(x)\hat{\psi}^{\dagger}_{B}(x)\hat{\psi}_{B}(x)\hat{\psi}_{B}(x), \nonumber \\
\hat{H}_{IB} &= {g_{IB}}\int dx~\hat{\psi}^{\dagger}_{I}(x) \hat{\psi}^{\dagger}_{B}(x) \hat{\psi}_{B}(x) \hat{\psi}_{I}(x). \label{Hamiltonian_IB}
\end{align}
The term $\textit{h}_{\sigma}(x) = -\frac{\hbar^{2}}{2 m_{\sigma}}\frac{\partial^{2}}{\partial x^{2}}+ V_{DW}(x)$ is the single-particle (SP) Hamiltonian for the $\sigma = I(B)$ species atoms and $V_{DW}(x) = a_{\sigma} (x^{2} - b_{\sigma}^{2})^{2}$ represents the DW. 
The two control parameters 
$a_{\sigma}$ and $b_{\sigma}$ 
adjust the central barrier height and 
the relative distance between the two wells, respectively. 
The field operator $ \hat{\psi}_{\sigma}^{\dagger}(x)$ [$\hat{\psi}_{\sigma}(x)$] 
creates (annihilates) a $\sigma$-species particle at position $x$~\cite{Pitaevskii_book}. 

Below, we rescale the MB Hamiltonian $\hat{H}$ in harmonic oscillator units which means that 
spatial and 
time scales are expressed in units of 
$\eta = \hbar  \omega$ and $\tau = 1/ \omega$, while the energy is rescaled in terms 
of $\xi = \sqrt{\hbar /m_{B} \omega}$. 
Furthermore, we assume zero or weak repulsive interactions among the bosons of the bath,  
which couple either repulsively or attractively with the impurity. 
Both intra- and interspecies interactions are modeled by short-range contact interaction potentials, being a legitimate approximation at ultra-cold temperatures, and are characterized by effective strengths $g_{BB}$ and $g_{IB}$, respectively~\cite{Feshbach_1,Feshbach_2,Feshbach_3,few_quench_3}. The latter can be controlled in the experiment through  
the $s$-wave scattering lengths via Feshbach or confinement-induced resonances \cite{Feshbach_1,Feshbach_2,Feshbach_3}.
Such a 1D mixture is experimentally accessible by imposing a tight transverse and a weak longitudinal confinement to a binary e.g., Bose-Fermi mixture consisting of two different isotopes of alkali atoms~\cite{mixture_exp_bf_1, mixture_exp_bf_2} or a Bose-Bose mixture composed by 
two different hyperfine states of the same isotope~\cite{mixture_exp_bb_1, mixture_exp_bb_2}. 
The DW potential can be 
easily realized by 
constructing a superlattice whose shells are double-wells~\cite{DW_exp_3, BJJ_2}. 

\begin{figure}
\centering
\includegraphics[width=0.5\textwidth]{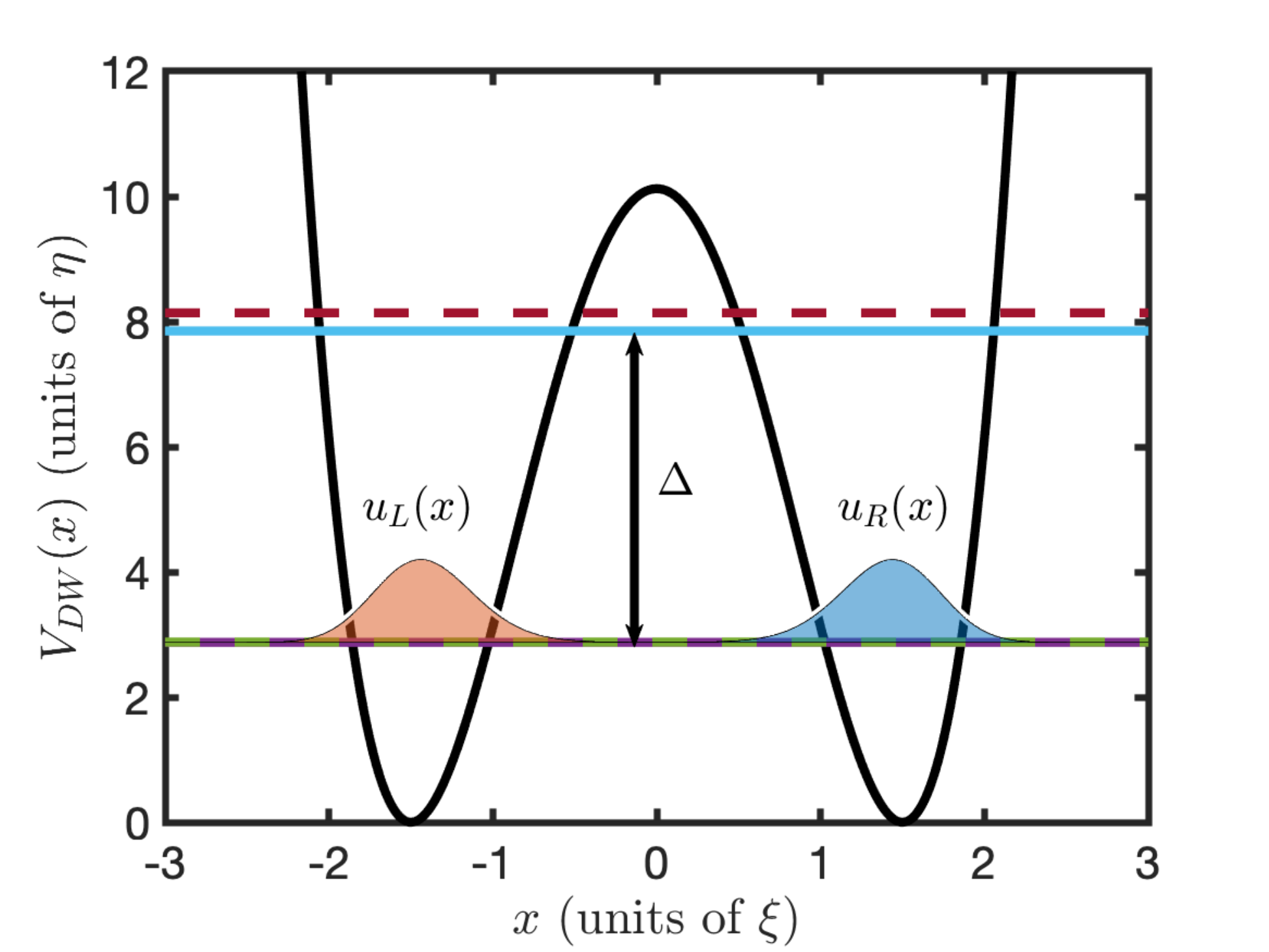}\hfill
\caption{(Color online) The external DW potential (black solid line), characterized by $a_{DW} = 2.0$, $b_{DW} = 1.5$, together with its energetically lowest four single-particle levels (horizontal lines). 
The solid (dashed) lines indicate the even (odd) parity states of the DW.
The energy gap between the lowest two energetic bands is marked by $\Delta$. The orange and blue shaded areas represent the lowest-band Wannier states $u_{L}(x)$ and $u_{R}(x)$ of the left and right wells, respectively.}
\label{DW_potential}
\end{figure}

For simplicity both species are trapped in the same DW geometry with $a_I = a_B = a_{DW}=2.0$ and $b_I = b_B = b_{DW}=1.5$. 
As a result, the low-lying SP spectrum of the DW forms a band-type doublet structure with $\Delta \approx 10^{3} \delta$. 
Here, $\Delta$ denotes the energy gap between the first and the second bands, while $\delta$ is the width of the first band. 
The DW potential is depicted in Fig.~\ref{DW_potential} with the black solid line and its lowest four SP energy levels are represented by the distinct horizontal lines.
Since the atoms of the bath are non-interacting or weakly repulsive (see above), it is possible 
to rely on  
the two-mode approximation
\begin{equation}
\hat{\psi}_{B}(x) = u_{L}(x) \hat{b}_{L} + u_{R}(x) \hat{b}_{R}, \label{2_mode_psi}
\end{equation}
with $u_{L,R}(x)$ being the Wannier states localized in the left and right well respectively as shown in 
Fig.~\ref{DW_potential}.  
This assumption leads to the low-energy effective Hamiltonian, i.e. the two-site Bose-Hubbard (BH) model, for the bath species
\begin{equation}
\hat{H}_{B} =  -J_{B} (\hat{b}^{\dagger}_{L}\hat{b}_{R} + \hat{b}^{\dagger}_{R} \hat{b}_{L}) + \frac{U_{B}}{2} \sum_{i = L,R} \hat{b}^{\dagger}_{i}\hat{b}^{\dagger}_{i}\hat{b}_{i}\hat{b}_{i}. \label{BH_model}
\end{equation}
The underlying hopping amplitude reads
\begin{align}
	J_{B} &= \int dx ~ u_{L}(x)\textit{h}_{B}(x)u_{R}(x),\label{BH_J} 
\end{align}
and the on-site repulsion energy is 
\begin{align}
	U_{B} &= g_{BB}\int dx ~ |u_{i}(x)|^{4} ~~~~~ (i = L,R).\label{BH_U}
\end{align}
Next, we designate the eigenstates of the single-species Hamiltonian $\hat{H}_{I}$ [$\hat{H}_{B}$] as $\{|\phi_{i}^{I}\rangle\}$ [$\{|\phi_{i}^{B}\rangle\}$], 
whilst the eigenstates of the bosonic SP Hamiltonian $h_{B}$ are $\{|\varphi^{B}_{i}\rangle \}$. 
Consequently, the Wannier states $|u_{L,R} \rangle$ are expressed in terms of $|\varphi^{B}_{0}\rangle$ and $|\varphi^{B}_{1}\rangle$ as $|u_{L,R}\rangle = \frac{1}{\sqrt{2}} [|\varphi^{B}_{0}\rangle \pm |\varphi^{B}_{1}\rangle]$. 
Also, each eigenstate of $\hat{H}_{\sigma}$ has  a definitive parity, such that the spatial parity symmetry of the DW is preserved, namely 
$|\phi_{i}^{\sigma}\rangle$ for $i = 0,2,4,\dots$ ($i = 1,3,5,\dots$) is of even (odd) parity.

In this work, we study the out-of-equilibrium  dynamics of the above highly particle imbalanced binary atomic mixture following a quench of the impurity-bath coupling strength from $g_{IB}=0$ to a finite value $g_{IB}\neq 0$. A particular focus is put on the microscopics of the 
time-evolved MB wave function $|\Psi(t) \rangle$ 
satisfying the Schr\"odinger equation $(i\hbar \partial / \partial t) | \Psi(t) \rangle = \hat{H} |\Psi(t) \rangle $.
The mixture is initially prepared in the ground state of the pre-quench Hamiltonian $|\Psi (0)\rangle = |\phi_{0}^{I}\rangle \otimes |\phi_{0}^{B}\rangle$, namely a product state of the GS of the DW for the impurity and 
the bath species Hamiltonians $\hat{H}_{I}$ and $\hat{H}_{B}$. 
To address the correlated dynamics of this system, we rely on the numerically exact diagonalization (ED) approach and obtain the time evolution in terms of an eigenstate decomposition. 
It enables us to obtain the time-dependent MB wave function of the system, while accounting for all emergent correlations.  

\section{Dynamical formation of two-fold fragmented MB state} \label{quench_weak}

To unravel the characteristics of the time-evolved MB wave function $|\Psi (t)\rangle$, we resort to a Schmidt decomposition 
\begin{equation}
	|\Psi (t)\rangle = \sum_{i=1}^{\mathcal{D}} \sqrt{\lambda_{i}(t)} ~|\psi_{i}^{I} (t) \rangle  |\psi_{i}^{B} (t)\rangle, \label{psi}
\end{equation}
where $ \lambda_{i} (t)$, being time-dependent real positive values, are the Schmidt numbers with $\lambda_1(t) > \lambda_2(t) >\dots$~\cite{Schmidt}. 
The normalization of the wave function is enforced by the constraint $\sum_{i} \lambda_i (t)= 1$. 
Moreover, $|\psi_{i}^{\sigma} (t)\rangle $ denotes the $i$th Schmidt orbital of the $\sigma$-species, obeying 
the orthogonality condition $\langle \psi_{i}^{\sigma} (t)| \psi_{j}^{\sigma} (t) \rangle = \delta_{i,j}$~\cite{Schmidt}. 
The upper bound $\mathcal{D}$ in the Schmidt decomposition is chosen from the fact that the maximal number of linear independent Schmidt states is $\min\{D_{I}, D_{B}\}$, with $D_{I}$ [$D_{B}$] being the Hilbert space dimension of the impurity [bath], see also the discussion below. 

Mathematically, $|\psi_{i}^{\sigma} (t)\rangle $ can always be expanded as a linear superposition of the $\hat{H}_{\sigma}$ eigenstates 
\begin{equation}
|\psi_{i}^{\sigma} (t)\rangle = \sum_{k=0}^{D_{\sigma}-1} C^{\sigma}_{i,k}(t) |\phi_{k}^{\sigma}\rangle, \label{SO_expansion}
\end{equation} 
with $C^{\sigma}_{i,k}(t) $ being the respective time-dependent expansion coefficients. 
The dimension, $D_{\sigma}$, of the $\sigma$-species Hilbert space refers to $D_{B} = N_B+1 = 101$ for the bosonic bath due to the employed two-mode approximation and for the impurity we truncate $D_{I}$ to an interaction-dependent finite large value which ensures   
the convergence of the employed one- and two-body observables. 
Before proceeding, it is important to note that the Schmidt numbers $\{ \lambda_{i} (t) \}$ quantitatively characterize the degree of entanglement between the two species~\cite{entanglement_1, Schmidt}. 
For instance, if multiple $\lambda_{i}(t)$ are non-vanishing 
a respective amount of product states $|\psi_{i}^{I} (t) \rangle  |\psi_{i}^{B} (t)\rangle$ 
contribute in the MB wave function
and hence the mixture is entangled. 
On the other hand, for the case of $ \lambda_{1}(t) = 1$ and $ \lambda_{i>1}(t) = 0$ the mixture is non-entangled, meaning that 
interspecies (but not necessarily intraspecies) correlations are suppressed in the course of the evolution.  
As we shall argue later on, in this limit the dynamics of the mixture is fully captured by the species mean-field (SMF) description, where the mutual impact of the species is merely an effective potential~\cite{ind_int_1,ind_int_2}.

\subsection{Dynamics of the Schmidt coefficients}

\begin{figure}
\centering
\includegraphics[width=0.5\textwidth]{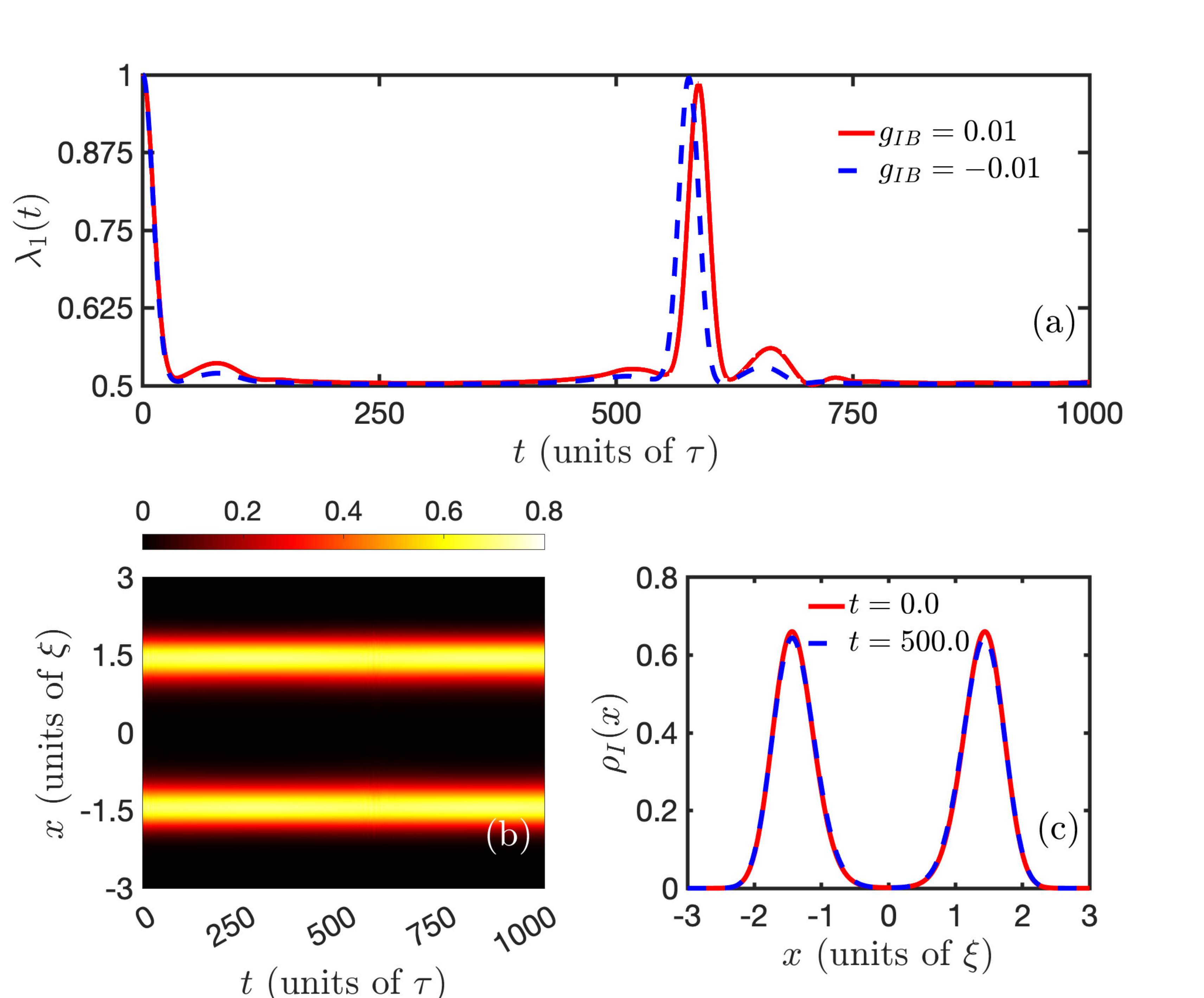}\hfill
\caption{(Color online) (a) Time-evolution of the first Schmidt number, $\lambda_{1}(t)$, for a quench from $g_{IB}=0$ to $g_{IB} = 0.01$ (red solid line) and towards $g_{IB} = -0.01$ (blue dashed line). 
The dynamical generation of a two-fold fragmented impurity state is evident since $\lambda_1(t)\approx \lambda_2(t) \approx 0.5$, and a revival behavior is observed for long evolution times. 
(b) The profile of the impurity's density $\rho^{I}_{1}(x,t)$ remains almost un-perturbed during the dynamics for a postquench impurity-bath coupling strength $g_{IB} = 0.01$. 
(c) Instantaneous density distributions (see legend) for the quench shown in 
panel (b).} 
\label{Snpop_I_gBB=000}
\end{figure}

Fig.~\ref{Snpop_I_gBB=000} (a) depicts the time-evolution of the first Schmidt number $\lambda_{1}(t)$ for fixed $g_{BB} = 0$ and for postquench impurity-bath couplings $g_{IB} = 0.01$ (red solid line) and $g_{IB} = -0.01$ (blue dashed line). 
Notice that the value of the second Schmidt number $\lambda_{2}(t)$ can be readily deduced via the relation $\lambda_{2}(t) \approx 1- \lambda_{1}(t)$, since for both quenches we have detected that the population of all higher-lying Schmidt numbers is negligible during the dynamics, namely $\lambda_i (t) < 10^{-4}$ for $i > 2$. 
Upon switching on the impurity-bath coupling, a "collapse-and-revival" behavior of $\lambda_{1}(t)$ is clearly observed for both repulsive and attractive postquench interactions. For instance, in the case of $g_{IB} = 0.01$ the first Schmidt number quickly decreases from $\lambda_{1}(0) = 1$ towards $\lambda_{1} = 0.502$ at $t = 38$ and remains almost as such 
for $t < 547$, before a pronounced revival occurs. 
This behavior of $\lambda_1(t)\approx \lambda_2(t)\approx 0.5$ is a direct manifestation of the \textit{dynamical formation of a two-fold fragmented MB state} in which the two product states $|\psi_{1}^{I} (t)\rangle |\psi_{1}^{B}(t) \rangle $ and $|\psi_{2}^{I} (t)\rangle |\psi_{2}^{B}(t) \rangle $ are equally populated 
for a long time interval. 
Hence, it holds that 
\begin{equation}
	|\Psi (t)\rangle \approx |\Psi_{F} (t)\rangle = \frac{1}{\sqrt{2}} \left[|\psi_{1}^{I} (t)\rangle |\psi_{1}^{B}(t) \rangle +|\psi_{2}^{I} (t) \rangle  |\psi_{2}^{B} (t) \rangle \right], \label{psi_type_I}
\end{equation} 
with $|\Psi_{F} (t)\rangle$ standing for the two-fold fragmented MB state. 
Further inspecting the corresponding Schmidt orbitals of the impurity, we find that 
\begin{equation}
|\psi_{1}^{I} (t) \rangle \approx |\phi_{0}^{I}\rangle, ~~~\textrm{and}~~~~~ |\psi_{2}^{I}(t) \rangle \approx |\phi_{1}^{I}\rangle. \label{NO_I_type_I}
\end{equation}
This observation immediately yields the SP occupations for the impurity as $\hat{n}_{0}^{sp} = \hat{n}_{1}^{sp} =1/2$ for $t \in [38,547]$. Here, $\hat{n}_{i}^{sp} = (\hat{a}_{i}^{I})^{\dagger}\hat{a}_{i}^{I}$ denotes the occupation number of the $i$th SP state $|\phi^{I}_{i}\rangle$ of the impurity, which relates to the band occupation as $\hat{n}_{1}^{Ib} = \hat{n}_{0}^{sp} + \hat{n}_{1}^{sp}$, $\hat{n}_{2}^{Ib} = \hat{n}_{2}^{sp} + \hat{n}_{3}^{sp}$, and so on. 
Let us note that Eq.~\eqref{NO_I_type_I} holds throughout the 
evolution times 
that we have examined, i.e., $t \in [0,1000]$. As a result, on the one hand, it indicates that the impurity motion is predominately restricted within the lowest band of the DW during the dynamics, i.e. $\hat{n}_{1}^{Ib}(t) = \hat{n}_{0}^{sp}(t) + \hat{n}_{1}^{sp}(t) \approx 1$. Hence, we observe that quenching $g_{IB}$ 
to the weak interaction regime triggers only intra-band excitations of the impurity. 
Due to the preserved parity symmetry of the mixture, the impurity's density profile $\rho^{I}_{1}(x,t) = \langle\Psi (t)|\hat{\psi}^{\dagger}_{I}(x) \hat{\psi}_{I}(x) |\Psi (t)\rangle / N_{I}$ remains nearly intact during evolution, i.e., $\rho^{I}_{1}(x,t) = \rho^{I}_{1}(x,0)$~\cite{polaron_BJJ} [see Figs.~\ref{Snpop_I_gBB=000}(b), (c)]. 
Recall that $\rho^{I}_{1}(x,t)$ estimates the probability of finding the impurity at position $x$ \cite{dmat_1, dmat_2}, and can be experimentally monitored through averaging a sample of single-shot measurements~\cite{cold_atom_rev}. 
On the other hand, it consequently results in an even (odd) parity Schmidt orbital $|\psi_{1}^{B} (t) \rangle$ ($|\psi_{2}^{B} (t)  \rangle$) for the bath species (see the discussion below). 

\begin{figure}
\centering
\includegraphics[width=0.5\textwidth]{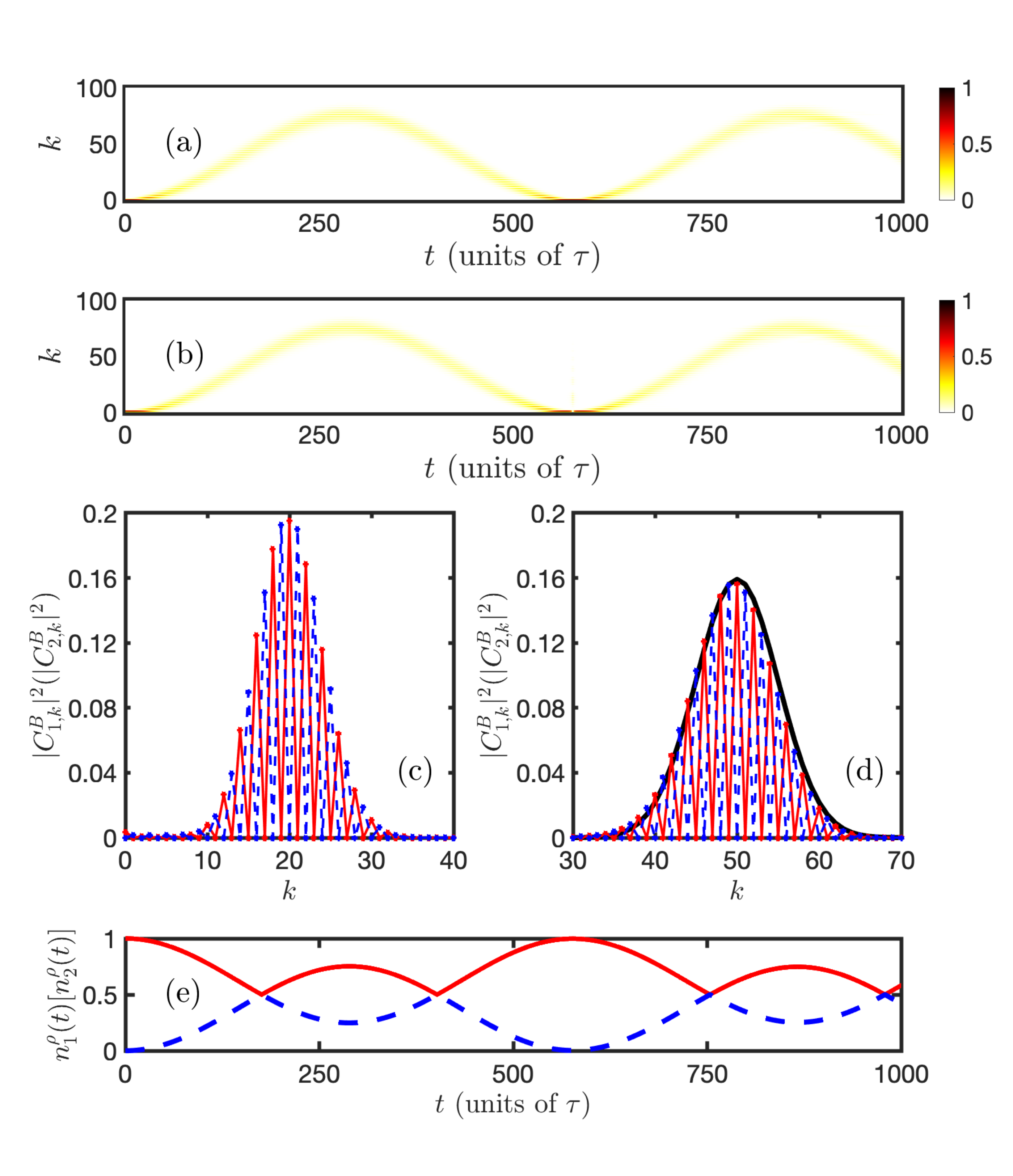}\hfill
\caption{(Color online) Evolution of the expansion coefficients (a) $|C^{B}_{1,k}(t)|^{2}$ and (b) $|C^{B}_{2,k}(t)|^{2}$ corresponding to the Schmidt orbitals $|\psi_{1}^{B} (t) \rangle$ and $|\psi_{2}^{B} (t) \rangle$ respectively for postquench impurity-bath interaction strength $g_{IB} = 0.01$. It is found that after the quench higher-lying excited states of the bath are populated.
Snapshots of the expansion coefficients $|C^{B}_{1,k}(t)|^{2}$ (red solid line) and $|C^{B}_{2,k}(t)|^{2}$ (blue dashed line) for (c) $t = 20$ and (d) $t = 175$. The binomial distribution [black solid line in (d)] is an imprint of the impurity-bath phase separation. 
(e) The corresponding time-evolution of the natural populations $n^{\rho}_{1}(t)$ (red solid line) and $n^{\rho}_{2}(t)$ (blue dashed line) for the quench to $g_{IB} = 0.01$ evincing the ensuing fragmentation process.}
\label{Fig_3}
\end{figure}

Having investigated the dynamics of the impurity, let us now turn to the bath species in order to gain deeper insights into the above-discussed dynamical formation of the MB two-fold fragmented state. 
We examine the behavior of the two Schmidt states $|\psi_{1}^{B} (t)\rangle$ and $|\psi_{2}^{B} (t)\rangle$ of the bath that are populated in the dynamics by utilizing  
Eq.~\eqref{SO_expansion}. 
The time-evolution of the corresponding expansion coefficients $C^{B}_{1,k}(t)$ and $C^{B}_{2,k}(t)$ for  
$g_{BB} = 0.0$ and $g_{IB} = 0.01$ is provided in Fig.~\ref{Fig_3}. 
Owing to the fact that the wave function is initially in a product form, i.e., $|\Psi (0)\rangle = |\phi_{0}^{I}\rangle \otimes |\phi_{0}^{B}\rangle$, we have $C^{B}_{1,0}(0) = 1$ and $C^{B}_{1,k}(0) = 0$ for $k>0$. Due to the non-interacting nature of the bath species, the GS wave function of $\hat{H}_{B}$ is simply equivalent to a mean-field state i.e. $\phi_{0}^{B} (x_1, \cdots, x_{N_{B}}) = \prod_{i=1}^{N_{B}} \varphi^{B}_{0} (x_{i})$, reflecting the fact that initially all the bosons are condensed into the SP state $|\varphi^{B}_{0}\rangle$~\cite{GPE} and therefore
$n_{1}^{\rho}(0) = 1$ and $n_{2}^{\rho}(0) = 0$. 
Here, $n_{i}^{\rho}(t)$ denote the natural populations obtained from a diagonalization of the reduced one-body density matrix of the bath species $\hat{\rho}_{1}^{B}(t) = \sum_{i=1}^{2} n^{\rho}_{i}(t) | \varphi_{i} (t)\rangle \langle \varphi_{i}(t)|$ \cite{dmat_1,dmat_2}, with $\{| \varphi_{i} (t)\rangle\} $ denoting the natural orbitals. 
Recall that the two-mode expansion of the field operator $\hat{\psi}(x)$ in Eq.~\eqref{2_mode_psi} leads to only two natural populations (natural orbitals) in the spectral decomposition. 
Physically, the natural population $n^{\rho}_{i}(t)$ denotes the probability for finding a single particle occupying the state $ | \varphi_{i}(t) \rangle$ at time $t$, after tracing out all other particles. 

For the cases where $n_{1}^{\rho}(t) < 1$, the motional excitations 
result in the emergence of quantum correlations in the dynamics, leading subsequently to depletion of the bath as can be testified by its natural populations  (see also below). 
Once switching on the impurity-bath coupling, we observe that the bath gradually populates its higher-lying excited states $|\phi_{k}^{B}\rangle$ for $k \gg 1$ [cf. Figs.~\ref{Fig_3}(a), (b)], accompanied by a slow depletion process [cf. Fig.~\ref{Fig_3}(e)]  
quantified by a decease (increase) of the natural population $n^{\rho}_{1}(t)$ [$n^{\rho}_{2}(t)$]. 
Before proceeding, it is important to point out that, at each time-instant, the coefficients $C^{B}_{1,k}(t)$ [$C^{B}_{2,k}(t)$] obtained from our ED simulations vanish exactly for the odd (even) parity eigenstates $|\phi_{k}^{B}\rangle$ as illustrated in Fig.~\ref{Fig_3}(c), (d). 
This observation is a direct consequence of the preserved parity symmetry imposed by the external DW of the mixture~\cite{polaron_BJJ}. 
Since the initial state $|\Psi (0)\rangle$ is of even parity, each product state $|\psi_{i}^{I} (t) \rangle |\psi_{i}^{B} (t) \rangle$ in Eq.~\eqref{psi} needs to exhibit an even parity 
and as a result the bosonic Schmidt orbital shares the same parity with the impurity. According to Eq.~\eqref{NO_I_type_I}, the Schmidt orbital $|\psi_{1}^{B} (t) \rangle$ ($|\psi_{2}^{B} (t) \rangle$) thus possesses an even (odd) parity~\cite{polaron_BJJ}. 

At $t = 175$, we observe that the bath reaches its maximal degree of depletion
characterized by $n_{1}^{\rho} = n_{2}^{\rho} = 0.5$, while both coefficients $|C^{B}_{1,k}|^{2}$ and $|C^{B}_{2,k}|^{2}$ follow a binomial distribution with their maximal values located at $k = 50$ and $k = 49$, respectively [see red solid and blue dashed line in Fig.~\ref{Fig_3}(d)]. It should be emphasized that the latter distribution indicates the emergence of a spatial two-body phase separation between the impurity and the bath~\cite{review1D_Mistakidis,Mistakidis_MB_pol} (see also the discussion below). Upon a basis transformation, the MB wave function at $t = 175$ can alternatively be written as 
\begin{equation}
	|\Psi_{C1}\rangle  = \frac{1}{\sqrt{2 N_{B}!}} \left[\hat{a}_{L}^{\dagger}(\hat{b}_{R}^{\dagger})^{N_{B}} + \hat{a}_{R}^{\dagger}(\hat{b}_{L}^{\dagger})^{N_{B}} \right] |0\rangle, \label{Psi_II_LR}
\end{equation}
with $\hat{a}_{L,R} = \frac{1}{\sqrt{2}} (\hat{a}_{0}^{I} \pm \hat{a}_{1}^{I})$ and $\hat{a}_{i}^{I}$ ($i=0,1$) referring to the annihilation operator acting on the $i$th SP state $|\phi_{i}^{I}\rangle$ of the impurity. Eq.~\eqref{Psi_II_LR} essentially represents a \textit{Schr\"odinger-cat state} being a superposition of two macroscopic MB states~\cite{entanglement_6,Streltsov_fragm_attract,Streltsov_fragm_barrier}. Each of them corresponds to a configuration where the impurity resides in one well while all the bosons are located in another well, thus manifesting the above-mentioned two-body phase separation between the two species. 
It should be emphasized that such a \textit{Schr\"odinger-cat state} is extremely sensitive 
meaning that an arbitrary small perturbation can lead to a collapse onto one of the two macroscopic configurations. 

\subsection{Two-body phase-separation at repulsive interspecies interactions}

\begin{figure*}
\centering
\includegraphics[width=\textwidth]{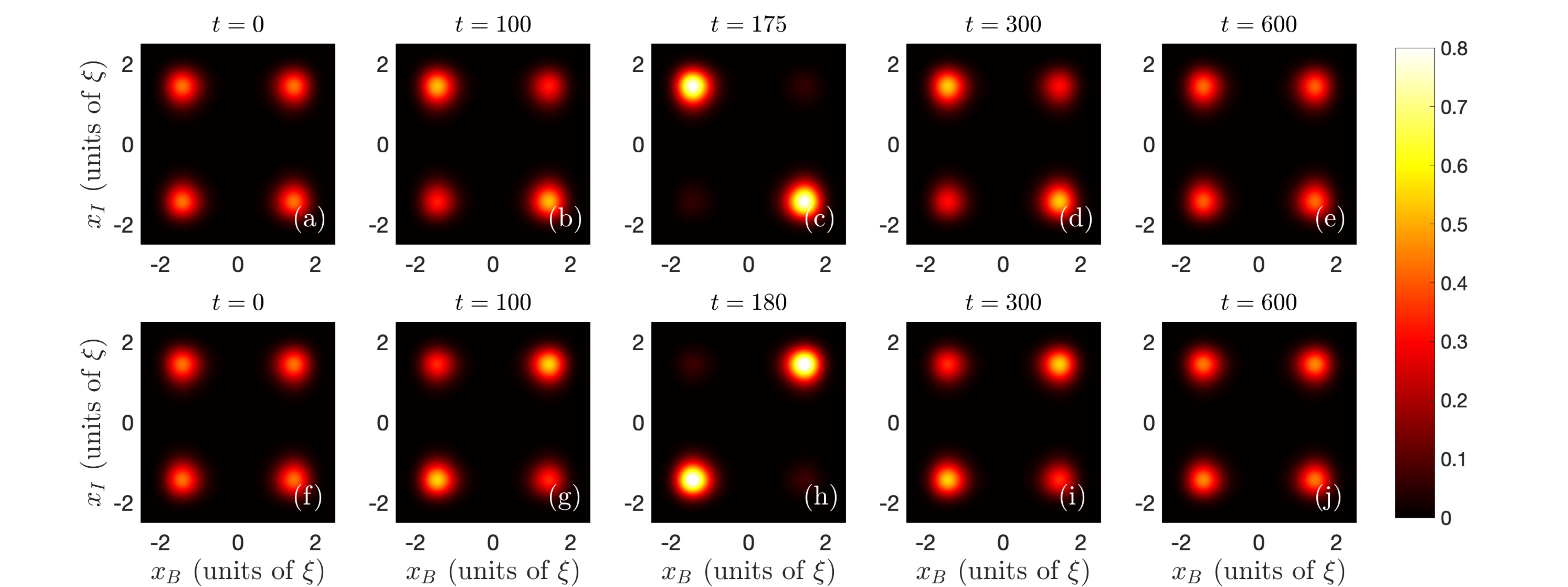}\hfill
\caption{(Color online) Impurity-bath two-body density distributions $\rho^{IB}_{2}(x_{I}, x_{B})$ upon considering a quench from $g_{IB}=0$ to $g_{IB} = 0.01$ (upper panels) and to $g_{IB} = -0.01$ (lower panels) at specific time-instants of the evolution (see legends). 
For $g_{IB}>0$, an antibunching behavior between the impurity and the bosons builds-up in the course of the evolution, see the increasing (decreasing) trend of the off-diagonal (diagonal), manifesting the tendency for two-body phase separation. However, in the case of $g_{IB}<0$ a clustering trend among the impurity and the bath bosons occurs as identified by the enhanced (decreased) population of the diagonal.}
\label{dmat2_gBB=000_gIB}
\end{figure*}

This two-body phase separation process can also be directly seen in the corresponding two-body impurity-bath density distribution $\rho^{IB}_{2}(x_{I}, x_{B}) = \langle\Psi|\hat{\psi}^{\dagger}_{I}(x_{I}) \hat{\psi}_{I}(x_{I}) \hat{\psi}^{\dagger}_{B}(x_{B}) \hat{\psi}_{B}(x_{B}) |\Psi\rangle/(N_{I}N_{B})$. 
This observable refers to the probability of detecting the impurity at position $x_I$ while one boson resides at location $x_B$. 
At $t = 0$ the two species are fully decoupled implying that both the impurity and the bosons can freely tunnel between the two wells. 
As a result, $\rho^{IB}_{2}(x_{I}, x_{B})$ exhibits four dominant peaks within the spatial regions ($x_{I}, x_{B}) = (\pm1.5, \pm1.5$), i.e., around the minima of the DW [cf. Fig.~\ref{DW_potential} and Fig.~\ref{dmat2_gBB=000_gIB}(a)]. 
However, a finite 
impurity-bath repulsion renders such a two-body density distribution energetically unfavorable. 
Consequently, the impurity and the bosons feature an 
antibunching as time evolves, a mechanism that can inferred by the  increase (decrease) of $\rho^{IB}_{2}(x_{I}, x_{B})$ values in the vicinity of the off-diagonal (diagonal) [cf. Fig.~\ref{dmat2_gBB=000_gIB} (b)]. 
As time evolves, e.g. at $t = 175$ the two species become fully anti-bunched, with the two-body probability distribution being solely occupied along its off-diagonal [see Fig.~\ref{dmat2_gBB=000_gIB} (c)]. 
The mixture, in turn, resides in a superposition of two equally-weighted configurations forming a \textit{Schrödinger-cat state}~\cite{Impurity_BH_2}, which manifests the dominant role of the impurity-bath entanglement. 
Recall that a similar two-body phase separation process has been recently demonstrated to occur in few-body ensembles and found to be related with the emergence of anti-ferromagnetic order~\cite{Murmann_few_fermions,Erdmann_phase_sep,Barfknecht_separation_few}.

\subsection{Clustering for attractive impurity-bath couplings}

It is worth mentioning that a similar dynamical formation of a two-fold fragmented MB state is also observed for 
impurity-bath interaction quenches towards the weakly attractive regime, see in particular the dynamics of the first Schmidt number $\lambda_{1}(t)$ for $g_{IB} = -0.01$ depicted in Fig.~\ref{Snpop_I_gBB=000}(a) by the blue dashed line and the dynamics of the impurity-bath two-body density presented in Fig.~\ref{dmat2_gBB=000_gIB}(f)-(j). 
Additionally, the validity of 
Eq.~\eqref{NO_I_type_I} holds equally for the attractive interaction case indicating that the impurity dynamics is dominated by intra-band excitation processes. 
However, in the $g_{IB}<0$ scenario, the configurations of the bath Schmidt states 
deviate from the ones of $g_{IB}>0$. 
In particular, for $t = 180$ the corresponding expansion coefficients $C^{B}_{1,k}(t)$ and $C^{B}_{2,k}(t)$ possess a negative sign as compared to the ones for $g_{IB} = 0.01$. 
For this reason, the underlying \textit{Schr\"odinger-cat state} becomes  
\begin{equation}
	|\Psi_{C2}\rangle  = \frac{1}{\sqrt{2 N_{B}!}} \left[\hat{a}_{L}^{\dagger}(\hat{b}_{L}^{\dagger})^{N_{B}} - \hat{a}_{R}^{\dagger}(\hat{b}_{R}^{\dagger})^{N_{B}} \right] |0\rangle, \label{Psi_II_RR}
\end{equation}
indicating a two-body clustering in space between the impurity and the bath, see also Fig.~\ref{dmat2_gBB=000_gIB}(h) where this two-body clustering tendency is clearly evident.

\section{Destruction of the two-fold fragmentation at strong interactions} \label{quench_strong}

\begin{figure}
\centering
\includegraphics[width=0.5\textwidth]{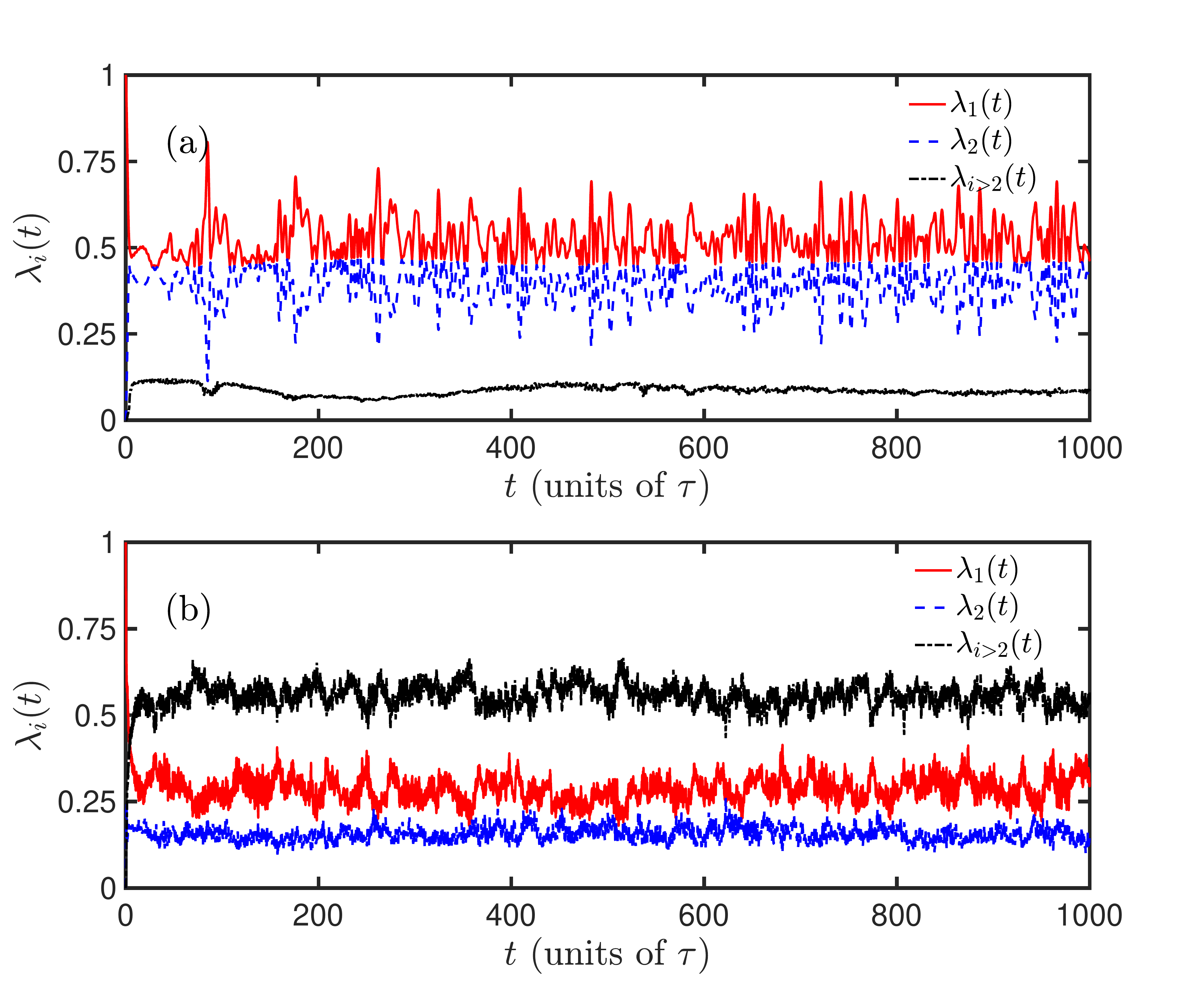}\hfill
\caption{(Color online) (a) Dynamics of the Schmidt coefficients $\lambda_1(t)$ (red solid line), $\lambda_2(t)$ (blue dashed line) as well as the total contribution of the remaining ones $\sum_{i>2} \lambda_{i}(t)$ (black dash-dotted line) following a quench from $g_{IB}=0$ towards $g_{IB} = 0.1$. 
(b) Same as (a) but for a postquench coupling $g_{IB} = 1.0$. 
In both cases higher than the first two Schmidt states are substantially occupied, signifying the inter-band excitation dynamics of the impurity and leading to the destruction of the two-fold fragmented MB state.}
\label{SN_gIB_large}
\end{figure}

Subsequently, we explore the impact of impurity-bath interaction quenches in the strongly coupled regime on the dynamical formation of a two-fold fragmented MB state.  
Without loss of generality, hereafter, we mainly consider the cases of repulsive postquench interaction strengths, while the scenario of relatively large attractive $g_{IB}<0$ will be mentioned for reasons of comparison. 
Fig.~\ref{SN_gIB_large} depicts the time-evolution of the Schmidt numbers for postquench couplings $g_{IB} = 0.1$ [Fig.~\ref{SN_gIB_large}(a)] and $g_{IB} = 1.0$ [Fig.~\ref{SN_gIB_large}(b)]. 
In sharp contrast to the weakly interacting regime where a "collapse-and-revival" behavior is observed in the largest two Schmidt numbers [see Fig.~\ref{Snpop_I_gBB=000}], increasing the impurity-bath coupling results in a rapid irregular oscillatory behavior for both $\lambda_{1}(t)$ and $\lambda_{2}(t)$, see Fig.~\ref{SN_gIB_large}. 

As a matter of fact, we conclude that within the strongly repulsive interspecies interaction regime 
the two-fold fragmented MB state can not be generated due to the non-negligible contribution of higher-lying Schmidt coefficients.   
Indeed, in the course of the evolution, 
there is a rapid growth of the  
Schmidt numbers $\lambda_{i}(t)$ with $i >2$ in both cases of $g_{IB} = 0.1$ and $g_{IB} = 1.0$. 
Recall that $\lambda_{i>2}(t)\approx 0$ in the weakly interacting regime, see Fig.~\ref{Snpop_I_gBB=000}. 
Importantly, for the case of $g_{IB} = 1.0$, we observe that the total population of the 
higher-lying Schmidt coefficients $\sum_{i>2} \lambda_{i}(t)$ becomes significantly greater than the contribution of both the first and the second Schmidt numbers already at the initial stages of the dynamics, e.g. for $t>10$ as shown in Fig.~\ref{SN_gIB_large}. 
Naturally, this observation 
reflects the fact that the participation of the respective 
higher-order product states $|\psi_{i}^{I} (t) \rangle  |\psi_{i}^{B} (t)\rangle$ ($i >2$) in the underlying MB wave function $|\Psi (t)\rangle $ is substantial.  

\begin{figure}
\centering
\includegraphics[width=0.5\textwidth]{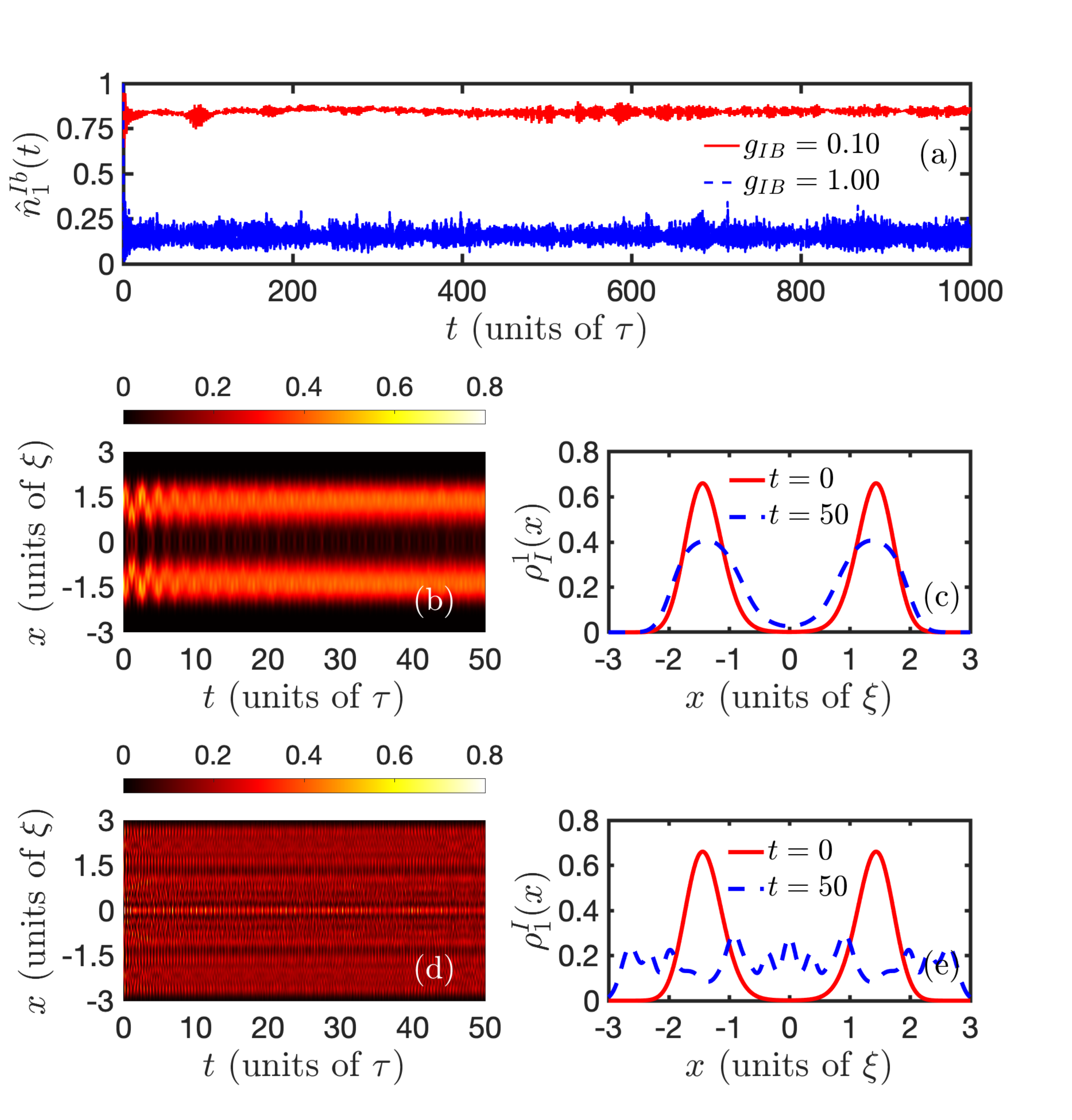}\hfill
\caption{(Color online) (a) Time-evolution of the first band occupation $\hat{n}_{1}^{Ib}(t)$ of the impurity for postquench interaction strength $g_{IB} = 0.1$ (red solid line) and for $g_{IB} = 1.0$ (blue dashed line). 
The decrease of $\hat{n}_{1}^{Ib}(t)$ marks the existence of inter-band excitation impurity processes. 
Spatiotemporal evolution of the impurity's density distribution, $\rho^{I}_{1}(x,t)$, after a quench to (b) $g_{IB} = 0.1$ showing intrawell breathing motion and (d) towards $g_{IB} = 1.0$ experiencing a multi-hump delocalized response. The latter designates the interband excitations processes of the impurity. 
(c) [(e)] Impurity density snapshots (see legends) taken from (b) [(d)].}
\label{band_gpopA_large_gIB}
\end{figure}

At this point, let us remark that the appreciable population of the higher-lying Schmidt orbitals alternatively implies the existence of inter-band excitation processes of the impurity as it was already argued for the ground state properties of the system~\cite{polaron_BJJ}. 
In fact, if the impurity is only restricted within the lowest-band of the DW, each Schmidt orbital of the impurity is a linear superposition of the SP states $|\phi_{0}^{I} \rangle$ and $|\phi_{1}^{I} \rangle$. 
To preserve orthogonality, i.e., $\langle \psi_{i}^{I}(t)|\psi_{j}^{I}(t)\rangle = 0$ for $i \neq j$, there are at most two linearly independent Schmidt orbitals populated in the underlying MB wave function $|\Psi (t)\rangle $. 
This property can be indeed confirmed by inspecting the time-evolution of the first band occupation $\hat{n}_{1}^{Ib}(t)$ of the impurity (see also the above discussion). 
As it can be seen in Fig.~\ref{band_gpopA_large_gIB} (a),  $\hat{n}_{1}^{Ib}(t)$ decreases rapidly 
at short evolution times featuring afterwards an irregular 
oscillatory behavior around an interaction-dependent 
average value, e.g. $\bar{n}_{1}^{Ib} = 0.82$ for $g_{IB} = 0.1$ and $\bar{n}_{1}^{Ib} = 0.15$ for $g_{IB} = 1.0$. 

The underlying inter-band excitation processes of the impurity in this strongly repulsive interaction regime are naturally also imprinted  
in its spatial density distribution $\rho^{I}_{1}(x,t)$. 
We again note that the density distribution of the bosonic bath is weakly perturbed throughout the time-evolution due to the employed two-mode approximation. 
However, the impurity's motion is strongly affected by the 
strong interspecies interactions, e.g., $g_{IB} = 0.1$, especially 
as compared to the above-described weakly interacting case [see Fig.~\ref{Snpop_I_gBB=000}(b)].  
Monitoring $\rho^{I}_{1}(x,t)$ in Fig.~\ref{band_gpopA_large_gIB}(b) we observe that it undergoes a complex 
oscillatory motion while remaining within the bath throughout the dynamics [see also Fig.~\ref{band_gpopA_large_gIB}(c), the snapshots of $\rho^{I}_{1}(x,t)$ for different times]. 
More specifically, it initially travels towards the center of the DW due to the impurity-bath repulsion and at $t = 2$ it is bounced back towards the left and right well, respectively. 
Overall, an intrawell breathing-type motion of the impurity takes place at short evolution times. 
Due to the interspecies energy exchange, in particular from the impurity to the bath~\cite{Koushik_driven_pol,Mistakidis_dissipative_pol,Mistakidis_rf_Bosepol}, 
this impurity breathing dynamics becomes quickly dissipative while, at longer timescales, 
$\rho^{I}_{1}(x,t)$ is restricted within each well and it is vanishing around 
the central barrier of the DW, see Fig.~\ref{band_gpopA_large_gIB}(b).

Quenching to stronger interspecies repulsions, e.g. $g_{IB}=1.0$, the impurity features a more rapid and irregular oscillatory response 
characterized by a multi-hump spatial density structure $\rho^{I}_{1}(x,t)$ as illustrated in Fig.~\ref{band_gpopA_large_gIB}(d) [see also Fig.~\ref{band_gpopA_large_gIB}(e) the snapshots of $\rho^{I}_{1}(x,t)$ for different times]. 
This multi-hump density configuration 
evinces that the impurity populates a superposition of higher-lying excited states of the DW, manifesting the dominant contribution of the impurity's inter-band excitations. 
Such excitation mechanisms of the impurity occurring in the strongly repulsive regime have been also observed for harmonically trapped impurities and they are related with the temporal orthogonality catastrophe phenomenon of the Bose polaron~\cite{Mistakidis_orth_cat,Mistakidis_MB_pol}. 
In our system the polaron behavior as quantified by the residue, i.e. the overlap between the MB state with vanishing impurity-bath  interactions and finite ones, is decaying but not completely suppressed due to the remaining partial overlap among the impurity and the bath.
Before proceeding, we remark 
that for 
interaction quenches towards the strongly attractive regime, the inter-band excitations of the impurity are significantly suppressed as compared to the cases for $g_{IB}>0$. 
Correspondingly, 
weak spatial distortions are observed in the 
time-evolution of the density distributions (results not shown for brevity). 

\begin{figure}
\centering
\includegraphics[width=0.5\textwidth]{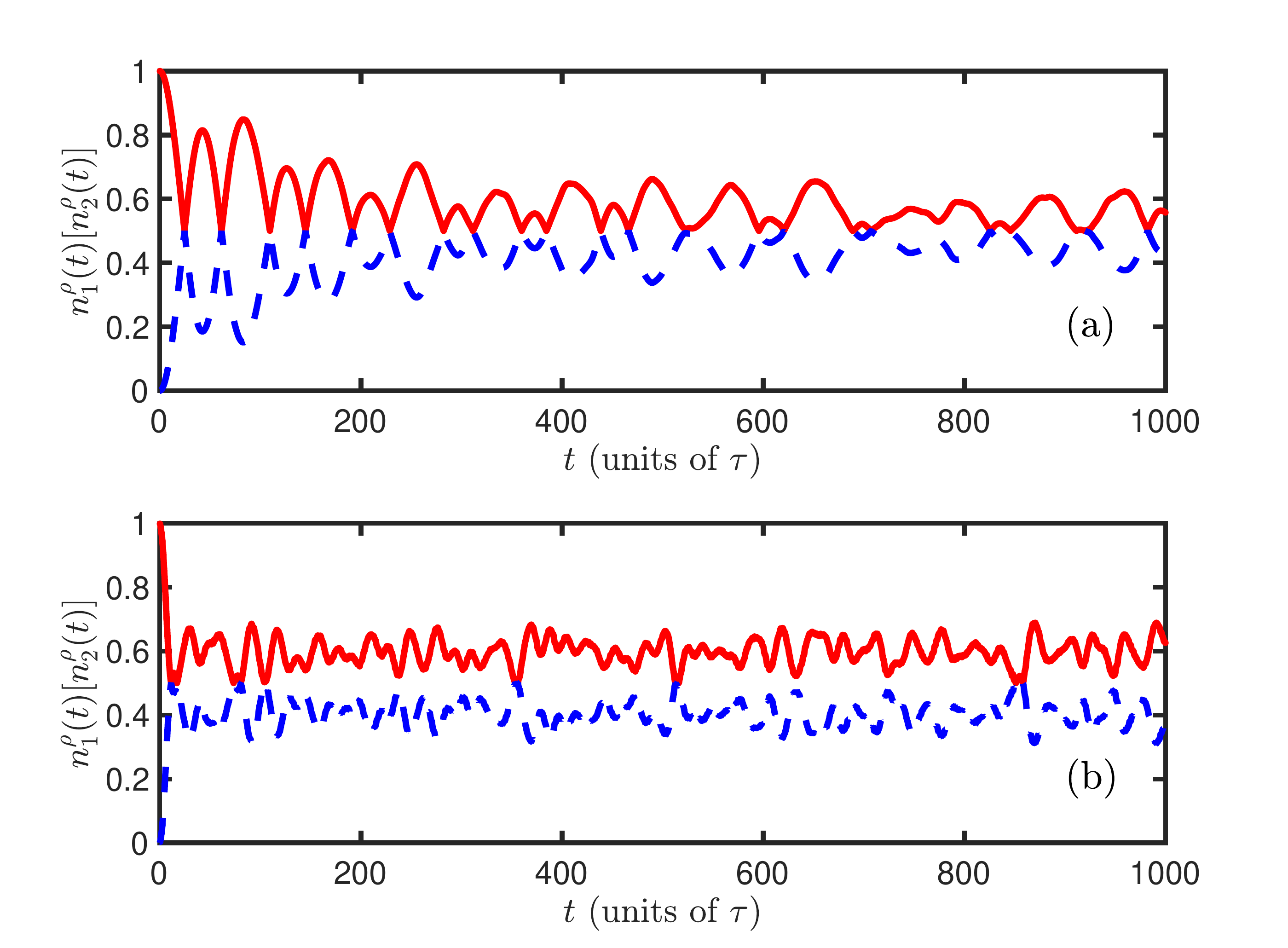}\hfill
\caption{(Color online) Temporal evolution of the natural populations $n_{1}^{\rho}(t)$ (red solid line) and $n_{2}^{\rho}(t)$ (blue dashed line) for postquench interaction $g_{IB} = 0.1$. 
(b) Same with (a) but for $g_{IB} = 1.0$. 
Dynamical fragmentation of the bath bosons is observed in both cases.}
\label{npop_B_large_gIB}
\end{figure}

As for the bath, albeit the fact that its density distribution is nearly unaltered during the evolution, we observe a substantial degree of dynamical fragmentation 
among the bosons. 
This process is signified by the rapid decrease (increase) of the corresponding natural population $n_{1}^{\rho}(t)$ ($n_{2}^{\rho}(t)$) in the course of the evolution. 
Indeed, as shown in Fig.~\ref{npop_B_large_gIB},  
$n_{1}^{\rho}(t)$ drops to the value $n_{1}^{\rho}=0.5$ 
at, for example, $t = 20$ for $g_{IB} = 0.1$ [Fig.~\ref{npop_B_large_gIB}(a)]  and $t =5$ for $g_{IB} = 1.0$ [Fig.~\ref{npop_B_large_gIB}(b)], respectively.
Importantly, in both cases $n_{1}^{\rho}(t)$ never acquires 
its original value $n_{1}^{\rho} = 1$ which is in contrast to the observations made within the 
weakly interacting regime [see also Fig.~\ref{Fig_3}].  
We also remark that in Ref.~\cite{BJJ_chaos}, it was demonstrated that such a dynamical behavior of the natural populations hints to the occurrence of quantum chaotic behavior of the bath species induced by the presence of the impurity. 
It manifests as an irregular motion for the bath species e.g., in its density population imbalance between the two wells as well as the existence of an effective Bose-Bose attraction.

\section{Quench dynamics with a weakly interacting bath} \label{interacting_bath}

\begin{figure}
\centering
\includegraphics[width=0.5\textwidth]{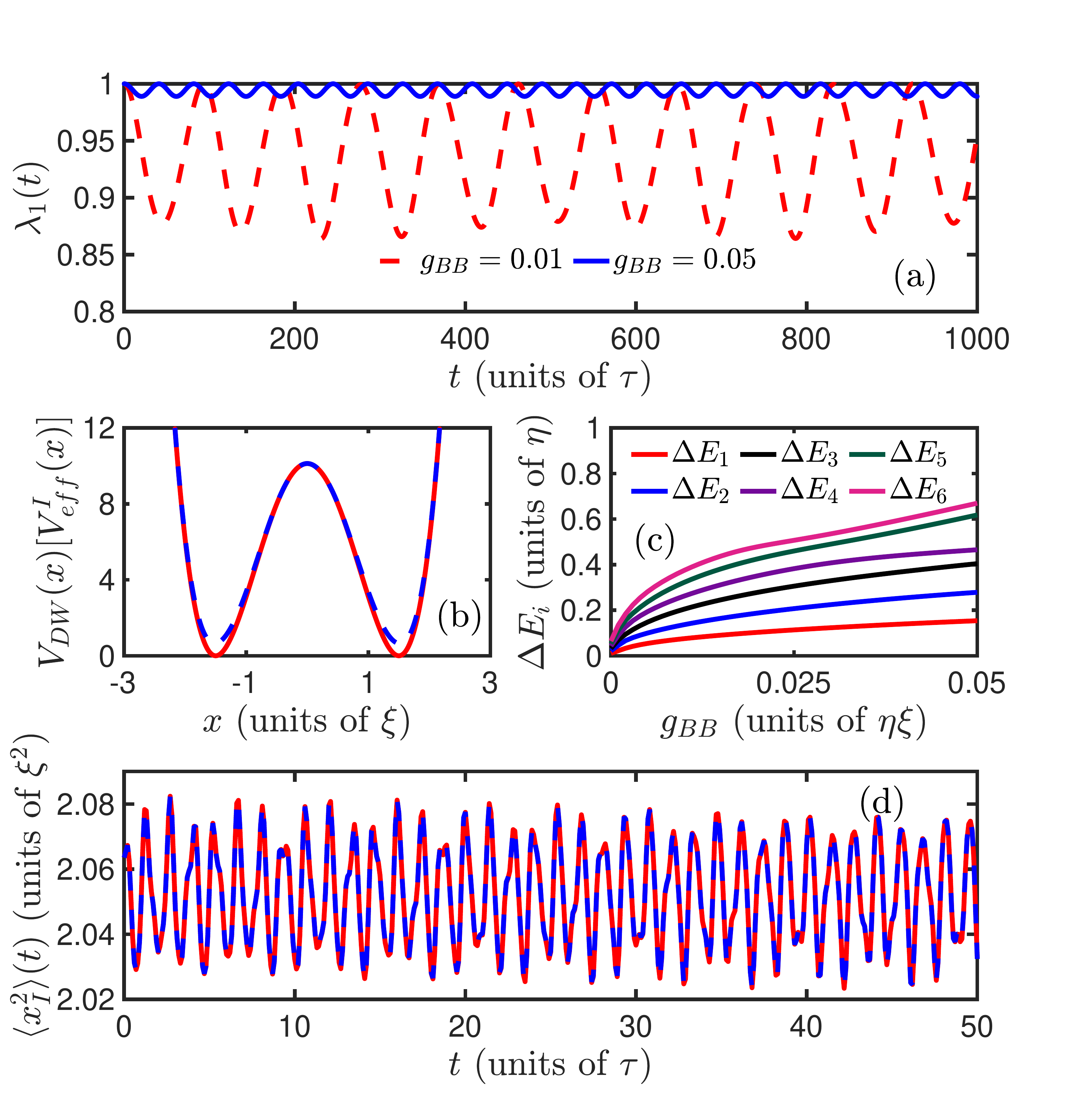}\hfill
\caption{(Color online) (a) Time-evolution of the first Schmidt number $\lambda_1(t)$ for the postquench interaction $g_{IB} = 0.01$ and for Bose-Bose repulsion $g_{BB} = 0.01$ (red dashed line) and $g_{BB} = 0.05$ (blue solid line). 
The small deviations of $\lambda_1(t)$ from unity imply the validity of the SMF approximation for the description of the MB wave function. 
(b) Spatial profile of the DW (red solid line) as well as the effective potential (blue dashed line) of the impurity corresponding to the case of $g_{IB} = 0.01$ and $g_{BB} = 0.05$. 
(c) Energy difference between the $i$th excited state and the GS of $\hat{H}$ denoted by $\Delta E_{i}$ as a function of $g_{BB}$ for fixed $g_{IB} = 0.01$. 
(d) Dynamics of the second moment of the impurity's position, $\langle x_I^2(t) \rangle$, obtained from the SMF approximation (blue dashed line) and from the MB ED simulation (red solid line). 
In both cases $g_{BB} = 0.05$ and the postquench impurity-bath coupling is $g_{IB} = 0.01$. 
The irregular oscillatory behavior of $\langle x_I^2(t) \rangle$ is an imprint of the impurity's breathing motion.}
\label{Fig_8}
\end{figure}

As a next step, we investigate how a weakly interacting bath affects the response and degree of fragmentation of the impurity. 
Without lost of generality, we set the Bose-Bose coupling strength to $g_{BB} = 0.01, 0.05$, which in turn results in the following ratio between the hopping amplitude and on-site interactions  $J_{B}/U_{B} = 4.78, 0.95$ [cf.~Eqs.~\eqref{BH_J} and \eqref{BH_U}]. 
The dynamics of the first Schmidt number $\lambda_{1}(t)$ is presented in Fig.~\ref{Fig_8}(a) for fixed $g_{IB} = 0.01$ and $g_{BB} = 0.01$ (red dashed line) or $g_{BB} = 0.05$ (blue solid line). 
In contrast to the case of a non-interacting bath, the presence of a finite repulsive $g_{BB}$ 
leads to a drastically different behavior of the Schmidt numbers during the dynamics. 
For example, when $g_{BB} = 0.01$, we observe that $\lambda_{1}(t)$ firstly decreases to the value of $\lambda_{1} = 0.88$ around $t = 50$ and then turns back to $\lambda_{1} \approx 1$ at $t = 98$ exhibiting such a persistent oscillation throughout the dynamics. 

Since the first Schmidt number lies close to unity 
for $g_{BB}>0$, the corresponding MB wave function can be well approximated as $|\Psi (t)\rangle = |\psi_{1}^{I} (t) \rangle  |\psi_{1}^{B} (t)\rangle$, i.e. it has a product form between the impurity and the bath Schmidt orbitals. 
Therefore, in this limit the mixture is fully captured by the SMF description~\cite{ind_int_1,polaron_BJJ} according to which the impurity experiences an effective potential consisting of the DW superimposed to a potential proportional to the bosonic density of the bath, namely $V_{\text{eff}}^{I}(x) = V_{DW}(x)  + g_{IB} ~ N_{B}\rho_{1}^{B}(x)$. 
The spatial profile of this effective potential is depicted in  Fig.~\ref{Fig_8}(b). 
Equipped with this knowledge, we note that the quench of the impurity-bath coupling leads to a sudden change of the DW zero-point energy [cf. Fig.~\ref{Fig_8}(b)]. 
After the quench, the impurity exhibits a breathing-type motion within the effective potential. 
This response is captured by the second moment of the impurity's position $\langle x_{I}^2(t) \rangle$  featuring an oscillatory behavior, see, e.g., Fig.~\ref{Fig_8}(c) for the case $g_{IB} = 0.01$ and $g_{BB} = 0.05$. 

On the other hand, due to the large particle imbalance among the species, the effective potential experienced by the bath deviates negligibly from the initial DW. 
For this reason, the corresponding Schmidt orbital $|\psi^{B}_{1}\rangle$ of the bath resembles the GS of the Hamiltonian $\hat{H}_{B}$, i.e., $|\psi^{B}_{1}\rangle \approx |\phi_{0}^{B}\rangle$. 
Physically, a larger $g_{BB}$ can result in relatively larger energy differences between the GS and the excited states of $\hat{H}$, see in particular Fig.~\ref{Fig_8}(c) in which $\Delta E_{i}$ denotes the energy difference between the $i$th excited state and the GS of $\hat{H}$. 
Thus, the bath is hardly excited for fixed interspecies interaction strength~\cite{polaron_BJJ}.

\section{Conclusions and Outlook} \label{Conclusions}

We have investigated the non-equilibrium correlated quantum dynamics of a single impurity immersed in a non-interacting bosonic environment being confined within an one-dimensional double-well potential. 
To track all the emergent system correlations we employ a numerically exact diagonalization approach. 

In particular, we focus on the microscopic features of the time-evolved many-body wave function of the strongly particle imbalanced mixture upon considering quenches of the impurity-bath coupling strength. To analyze the obtained time-dependent MB state, we rely on a Schmidt decomposition in which the corresponding Schmidt numbers directly quantify the number of configurations that are macroscopically populated. 
For weak impurity-bath postquench interactions, we observe the dynamical formation of a persistent two-fold fragmented MB state where two almost equally weighted configurations contribute. 
The generation of this fragmented MB state is identified  by the "collapse-and-revival" behavior of the first two Schmidt numbers and it originates from the intra-band excitation processes of the impurity. 
It is also imprinted in the 
interspecies two-body density distributions which evinces  
a two-body phase separation (clustering) process among the two species for repulsive (attractive) interactions. 

Next, we unravel the impact of strong postquench  
interspecies interaction strengths on the formation of the two-fold fragmented state.  
It turns out that the impurity undergoes additional inter-band excitation processes with higher-lying 
Schmidt states having a substantial contribution.  
As a result, it destructs the dynamical generation of the two-fold fragmented MB state. 
A similar effect is observed when considering a finite Bose-Bose repulsion in the bath. 

Possible future investigations include the impact of a few 
impurities and/or the bare Bose-Bose repulsion on the MB excitation dynamics and the possible formation of few-body impurity clusters. 
Another interesting perspective is to study the correlated dynamics of the impurity setting 
but in the presence of long-range e.g. dipolar interactions~\cite{Ardila_dipol,Scheiermann_dipol}. Here, the impact of higher-band excitations of the impurity or a beyond the two-site Bose-Hubbard description for the bosonic component would be worthwhile to pursue.

\begin{acknowledgments}
This work has been funded by the Deutsche Forschungsgemeinschaft (DFG, German Research Foundation) - SFB 925 - project 170620586. S.I.M. gratefully acknowledges financial support 
from the NSF through a grant for ITAMP at Harvard University. 
\end{acknowledgments}


\begin{thebibliography}{10}

\bibitem{Cazalilla} M.A. Cazalilla, R. Citro, T. Giamarchi, E. Orignac, and M. Rigol, Rev. Mod. Phys. \textbf{83}, 1405 (2011).
\bibitem{Sowinski} T. Sowiński, and M.Á. García-March, Rep. Progr. Phys. \textbf{82}, 104401 (2019). 
\bibitem{review1D_Mistakidis} S. I. Mistakidis, A. G. Volosniev, R. E. Barfknecht, T. Fogarty, T. Busch, A. Foerster, P. Schmelcher, and N. T. Zinner, arXiv:\textbf{2202.11071} (2022).

\bibitem {cold_atom_rev} W. D. Phillips, Rev. Mod. Phys. {\bf70}, 721 (1998); I. Bloch, J. Dalibard, and W. Zwerger, Rev. Mod. Phys. {\bf 80}, 885 (2008).

\bibitem {Wenz} A.N. Wenz, G. Z\"urn, S. Murmann, I. Brouzos, T. Lompe, and S. Jochim, Science \textbf{342}, 457 (2013). 

\bibitem {Zurn} G. Z\"urn, F. Serwane, T. Lompe, A.N. Wenz, M.G. Ries, J.E. Bohn, and S. Jochim, Phys. Rev. Lett. \textbf{108}, 075303 (2012).

\bibitem{Feshbach_1} M. Olshanii, Phys. Rev. Lett. {\bf 81}, 938 (1998).
\bibitem{Feshbach_2} C. Chin, R. Grimm, P. Julienne, and E. Tiesinga, Rev. Mod. Phys. {\bf 82}, 1225 (2010).
\bibitem{Feshbach_3} T. K\"ohler, K. G\'oral, and P. S. Julienne, Rev. Mod. Phys. {\bf 78}, 1311 (2006).

\bibitem {BH_exp_1} M. Greiner, O. Mandel, T. Esslinger, T. W. H\"ansch, and I. Bloch, Nature (London) {\bf415}, 39 (2002).
\bibitem {BH_exp_2} I. B. Spielman, W. D. Phillips, and J. V. Porto, Phys. Rev. Lett. {\bf98}, 080404 (2007).
\bibitem {BH_exp_3} T. St\"oferle, H. Moritz, C. Schori, M. K\"ohl, and T. Esslinger, Phys. Rev. Lett. {\bf92}, 130403 (2004).

\bibitem{DW_exp_1} M. R. Andrews, C. G. Townsend, H.-J. Miesner, D. S. Durfee, D. M. Kurn, and W. Ketterle, Science {\bf275}, 637 (1997).
\bibitem{DW_exp_2} A. Rohrl, M. Naraschewski, A. Schenzle, and H. Wallis, Phys. Rev. Lett. {\bf78}, 4143 (1997).
\bibitem{DW_exp_3} M. Albiez, R. Gati, J. F\"olling, S. Hunsmann, M. Cristiani, and M. K. Oberthaler, Phys. Rev. Lett. {\bf95}, 010402 (2005).

\bibitem{BJJ_1} B. D. Josephson, Phys. Lett. {\bf1}, 251 (1962).
\bibitem{BJJ_2} R. Gati and M. K. Oberthaler, J. Phys. B: At. Mol. Opt. Phys. {\bf40} R61 (2007).


\bibitem{BJJ_Rabi_1} A. Smerzi, S. Fantoni, S. Giovanazzi, and S. R. Shenoy, Phys. Rev. Lett. {\bf79}, 4950 (1997).
\bibitem{BJJ_Rabi_2} S. Raghavan, A. Smerzi, S. Fantoni, and S. R. Shenoy, Phys. Rev. A {\bf59}, 620 (1999).
\bibitem{BJJ_Rabi_3} G. J. Milburn, J. Corney, E. M. Wright, and D. F. Walls, Phys. Rev. A {\bf55}, 4318 (1997).

\bibitem{BJJ_Squeeze_1} J. Est\`eve, C. Gross, A. Weller, S. Giovanazzi, and M. K. Oberthaler, Nature (London) {\bf455}, 1216 (2008).
\bibitem{BJJ_Squeeze_2} B. Juli\'a-D\'iaz, T. Zibold, M. K. Oberthaler, M. Mel\'e-Messeguer, J. Martorell, and A. Polls, Phys. Rev. A {\bf86}, 023615  (2012).

\bibitem{BJJ_Few_1}  S. Z\"ollner, H.-D. Meyer, and P. Schmelcher, Phys. Rev. Lett. {\bf100}, 040401 (2008).
\bibitem{BJJ_Few_2} B. Chatterjee, I. Brouzos, S. Z\"ollner, and P. Schmelcher, Phys. Rev. A {\bf82}, 043619 (2010).
\bibitem{BJJ_Few_3} S. Z\"ollner, H.-D. Meyer, and P. Schmelcher, Phys. Rev. A {\bf78}, 013621 (2008).
\bibitem{BJJ_Few_4} S. Z\"ollner, H.-D. Meyer, and P. Schmelcher, Phys. Rev. A {\bf74}, 063611 (2006).
\bibitem{BJJ_Few_5} S. Z\"ollner, H.-D. Meyer, and P. Schmelcher, Phys. Rev. A {\bf74}, 053612 (2006).

\bibitem{Bose_polaron_1} J. Catani, G. Lamporesi, D. Naik, M. Gring, M. Inguscio, F. Minardi, A. Kantian, and T. Giamarchi, Phys. Rev. A {\bf 85}, 023623 (2012).
\bibitem{Bose_polaron_2} T. Fukuhara, A. Kantian, M. Endres, M. Cheneau, P. Schaulß, S. Hild, D. Bellem, U. Schollw\"ock, T. Giamarchi, C. Gross, I. Bloch, and S. Kuhr, Nat. Phys. {\bf9}, 235 (2013).
\bibitem{Fermi_polaron_1} C. Kohstall, M. Zaccanti, M. Jag, A. Trenkwalder, P. Massignan, G.M. Bruun, F. Schreck, and R. Grimm, Nature {\bf485}, 615 (2012).
\bibitem{Fermi_polaron_2} M. Koschorreck, D. Pertot, E. Vogt, B. Fr\"ohlich, M. Feld, and M. K\"ohl, Nature {\bf485}, 619 (2012).
\bibitem{Fermi_polaron_3} F. Scazza, G. Valtolina, P. Massignan, A. Recati, A. Amico, A. Burchianti, C. Fort, M. Inguscio, M. Zaccanti, and G. Roati, Phys. Rev. Lett. {\bf 118}, 083602 (2017).

\bibitem{polaron_conmat_1} L. D. Landau, Phys. Z. Sowjetunion {\bf3}, 644 (1933).
\bibitem{polaron_conmat_2} L. D. Landau and S. I. Pekar, Zh. Eksp. Teor. Fiz. {\bf18}, 419 (1948).
\bibitem{polaron_conmat_3} R. P. Feynman, Phys. Rev. {\bf97}, 660 (1955).
\bibitem{polaron_conmat_4} P. W. Anderson, Phys. Rev. Lett. {\bf18}, 1049 (1967).

\bibitem {Massignan_rev} P. Massignan, M. Zaccanti, and G.M. Bruun, Rep. Progr. Phys. \textbf{77}, 034401 (2014). 
\bibitem {Schmidt_rev} R. Schmidt, M. Knap, D.A. Ivanov, J.S. You, M. Cetina, and E. Demler, Rep. Progr. Phys. \textbf{81}, 024401 (2018).

\bibitem {Johnson_lat} T. H. Johnson, S. R. Clark, M. Bruderer, and D. Jaksch,  Phys. Rev. A \textbf{84}, 023617 (2011)
\bibitem {Palzer_lat} S. Palzer, C. Zipkes, C. Sias, and M. K{\"o}hl,  Phys. Rev. Lett. \textbf{103}, 150601 (2009). 
\bibitem {Bruderer_lat} M. Bruderer, A. Klein, S.R. Clark, and D. Jaksch, New J. Phys. \textbf{10}, 033015 (2008).
\bibitem {Theel_DW} F. Theel, K. Keiler, S. I. Mistakidis, and P. Schmelcher,  Phys. Rev. Research \textbf{3}, 023068 (2021). 

\bibitem {Dehkharghani_ind} A. S. Dehkharghani, A. G. Volosniev, and N. T. Zinner, Phys. Rev. Lett. \textbf{121}, 080405 (2018). 
\bibitem{ind_int_1} J. Chen, J. M. Schurer, and P. Schmelcher, Phys. Rev. Lett. {\bf121}, 043401 (2018). 
\bibitem {Mistakidis_ind} S. I. Mistakidis, A. G. Volosniev, and P. Schmelcher,  Phys. Rev. Research, \textbf{2}, 023154 (2020). 
\bibitem {Mukherjee_ind} K. Mukherjee, S. I. Mistakidis, S. Majumder, and P. Schmelcher, Phys. Rev. A \textbf{102}, 053317 (2020). 

\bibitem {Naidon} P. Naidon, J. Phys. Soc. Japan \textbf{87}, 043002 (2018). 
\bibitem {Alhyder} R. Alhyder, X. Leyronas, and F. Chevy, Phys. Rev. A \textbf{102}, 033322 (2020). 
\bibitem {Will} M. Will, G. E. Astrakharchik, and M. Fleischhauer,  Phys. Rev. Lett. \textbf{127}, 103401 (2021). 
\bibitem {Camacho_bipol} A. Camacho-Guardian, L. P. Ardila, T. Pohl, and G. M. Bruun, Phys. Rev. Lett. \textbf{121}, 013401 (2018). 
\bibitem {Brauneis_bound} F. Brauneis, T. G. Backert, S. I. Mistakidis, M. Lemeshko, H. W. Hammer, and A. G. Volosniev, New J. Phys. \textbf{24}, 063036 (2022). 

\bibitem {Yang_sol} M. Yang, M. {\v{C}}ufar, E. Pahl, and J. Brand, Cond. Matter \textbf{7}, 15 (2022). 
\bibitem {Koutentakis_sol} G. M. Koutentakis, S. I. Mistakidis, and P. Schmelcher, Atoms \textbf{10}, 3 (2021).  
\bibitem {Tajima_shock} H. Tajima, J. Takahashi, E. Nakano, and K. Iida, Phys. Rev. A \textbf{102}, 051302 (2020). 

\bibitem {Lausch_relaxation} T. Lausch, A. Widera, and M. Fleischhauer, Phys. Rev. A \textbf{97}, 023621 (2018).
\bibitem {Boyanovsky_dissipative_pol} D. Boyanovsky, D. Jasnow, X. L. Wu, and R. C. Coalson, Phys. Rev. A \textbf{100}, 043617 (2019). 
\bibitem {Mistakidis_dissipative_pol} S. I. Mistakidis, F. Grusdt, G. M. Koutentakis, and P. Schmelcher,  New J. Phys. \textbf{21}, 103026 (2019). 

\bibitem{entanglement_1} L. Amico, R. Fazio, A. Osterloh, and V. Vedral, Rev. Mod. Phys. {\bf80}, 517 (2008).

\bibitem{entanglement_2} W. Wang, V. Penna, and B. Capogrosso-Sansone, New J. Phys. {\bf18} 063002 (2016).
\bibitem{entanglement_3} A. Lukin, M. Rispoli, R. Schittko, M. E. Tai, A. M. Kaufman, S. Choi, V. Khemani, J. L\'eonard, and M. Greiner, Science {\bf364}, 256 (2019).
\bibitem{entanglement_4} H. Strobel, W. Muessel, D. Linnemann, T. Zibold, D. B. Hume, L. Pezze, A. Smerzi, and M. K. Oberthaler, Science {\bf345}, 424 (2014).
\bibitem{entanglement_5} A. Okopiska, J. Phys.: Conf. Ser. {\bf213}, 012004 (2010).

\bibitem{Impurity_BH_1} M. Rinck and C. Bruder, Phys. Rev. A {\bf83}, 023608 (2011).
\bibitem{Impurity_BH_2} F. Mulansky, J. Mumford, and D. H. J. O'Dell, Phys. Rev. A {\bf84}, 063602 (2011).
\bibitem{Impurity_BH_3} J. Mumford and D. H. J. O'Dell, Phys. Rev. A {\bf90}, 063617 (2014).
\bibitem{Impurity_BH_4} J. Mumford, J. Larson, and D. H. J. O'Dell, Phys. Rev. A {\bf89}, 023620 (2014).
\bibitem{Impurity_BH_5} J. Mumford, W. Kirkby, and D. H. J. O'Dell, J. Phys. B: At. Mol. Opt. Phys. {\bf53}, 145301 (2020).
\bibitem{Impurity_BH_6} J Mumford \textit{et al}., J. Phys. B: At. Mol. Opt. Phys. {\bf53}, 145301 (2020). 

\bibitem{Pitaevskii_book} L. Pitaevskii, and S. Stringari, Bose-Einstein condensation and superfluidity (Vol. 164). Oxford University Press (2016). 
\bibitem{few_quench_3} A. C. Pflanzer, S. Z\"ollner, and P. Schmelcher, J. Phys. B: At., Mol., Opt. Phys. {\bf42}, 231002 (2009).

\bibitem{mixture_exp_bf_1}  A. G. Truscott, K. E. Strecker, W. I. McAlexander, G. B. Partridge, and R. G. Hulet, Science {\bf291}, 2570 (2001).
\bibitem{mixture_exp_bf_2}  F. Schreck, L. Khaykovich, K. L. Corwin, G. Ferrari, T. Bourdel, J. Cubizolles, and C. Salomon, Phys. Rev. Lett. {\bf 87}, 080403 (2001). 
\bibitem{mixture_exp_bb_1} D. S. Hall, M. R. Matthews, J. R. Ensher, C. E. Wieman, and E. A. Cornell, Phys. Rev. Lett. {\bf81}, 1539 (1998).
\bibitem{mixture_exp_bb_2} S. Tojo, Y. Taguchi, Y. Masuyama, T. Hayashi, H. Saito, and T. Hirano, Phys. Rev. A {\bf 82}, 033609 (2010).

\bibitem {Schmidt}  A. Pathak, \textit{ Elements of Quantum Computation and Quantum Communication}, (Taylor \& Francis, 2013).

\bibitem{ind_int_2} J. Chen, J. M. Schurer, and P. Schmelcher, Phys. Rev. A {\bf 98}, 023602 (2018).
\bibitem{polaron_BJJ} J. Chen, S.I. Mistakidis, and P. Schmelcher, New J. Phys. {\bf24} 033004 (2022).

\bibitem {dmat_1} O. Penrose and L. Onsager, Phys. Rev. {\bf104}, 576 (1956).
\bibitem {dmat_2} K. Sakmann, A. I. Streltsov, O. E. Alon, and L. S. Cederbaum, Phys. Rev. A {\bf78}, 023615 (2008).
\bibitem {GPE} C. J. Pethick and H. Smith, \textit{Bose-Einstein Condensation in Dilute Gases}, (Cambridge University Press, New York, 2008).

\bibitem {Mistakidis_MB_pol} S. I. Mistakidis, G. M. Koutentakis, G. C. Katsimiga, T. Busch, and P. Schmelcher, New J. Phys. \textbf{22}, 043007 (2020). 

\bibitem{entanglement_6} G. C. Katsimiga, G. M. Koutentakis, S. I. Mistakidis, P. G. Kevrekidis, and P. Schmelcher, New J. Phys. \textbf{19}, 073004 (2017). 
\bibitem {Streltsov_fragm_attract} A. I. Streltsov, O. E. Alon, and L. S. Cederbaum, Phys. Rev. Lett. \textbf{100}, 130401 (2008). 
\bibitem {Streltsov_fragm_barrier} A. I. Streltsov, O. E. Alon, and L. S. Cederbaum, Phys. Rev. Lett. \textbf{99}, 030402 (2007). 

\bibitem {Murmann_few_fermions} S. Murmann, F. Deuretzbacher, G. Z\"urn, J. Bjerlin, S. M. Reimann, L. Santos, T. Lompe, and S. Jochim, Phys. Rev. Lett. \textbf{115}, 215301 (2015). 
\bibitem {Erdmann_phase_sep} J. Erdmann, S. I. Mistakidis, and P. Schmelcher, Phys. Rev. A \textbf{99}, 013605 (2019). 
\bibitem {Barfknecht_separation_few} R. E. Barfknecht, A. Foerster, and N. T. Zinner, Scientific Rep. \textbf{9}, 1-11 (2019). 

\bibitem {Koushik_driven_pol} K. Mukherjee, S. I. Mistakidis, S. Majumder, and P. Schmelcher, Phys. Rev. A \textbf{101}, 023615 (2020). 
\bibitem {Mistakidis_rf_Bosepol} S. I. Mistakidis, G. M. Koutentakis, F. Grusdt, H. R. Sadeghpour, and P. Schmelcher, New J. Phys. \textbf{23}, 043051 (2021). 

\bibitem {Mistakidis_orth_cat}  S. I. Mistakidis, G. C. Katsimiga, G. M. Koutentakis, T. Busch, and P. Schmelcher, Phys. Rev. Lett. \textbf{122}, 183001 (2019). 

\bibitem {BJJ_chaos}  J. Chen, K. Keiler, G.X. long, and P. Schmelcher, Phys. Rev. A {\bf104}, 033315 (2021).

\bibitem {Ardila_dipol} L. P. Ardila, and T. Pohl, Ground-state properties of dipolar Bose polarons. J. Phys. B: At. Mol. and Opt. Phys. \textbf{52}, 015004 (2018). 
\bibitem {Scheiermann_dipol} D. Scheiermann, L. A. Ardila, T. Bland, R. N. Bisset, and L. Santos, arXiv:\textbf{2202.08259} (2022).





\end{thebibliography}
\end{document}